\documentclass[10pt]{article}
\usepackage{amssymb}
\usepackage{amsfonts}
\usepackage{amsmath}
\usepackage{hyperref}

\makeatletter \@addtoreset{equation}{section} \makeatother

\newtheorem{theorem}{Theorem}
\newtheorem{lemma}{Lemma}

\newtheorem{remark}{Remark}

\begin{document}

\title{Fluctuations of the eigenvalue number  in the  fixed interval  for  $\beta$-models
with $\beta=1,2,4$}
\author{ M. Shcherbina
\\
\\
Institute for Low Temperature Physics, Ukr. Ac. Sci \\
 47 Lenin ave, 61135 Kharkov, Ukraine}
\date{}

\maketitle
This paper is dedicated to Prof. Brunello Tirozzi on the occasion of his 70th birthday.

\begin{abstract} We study the fluctuation of the eigenvalue number of any fixed interval $\Delta=[a,b]$ inside the spectrum
for $\beta$- ensembles of random matrices in the case $\beta=1,2,4$. We assume that the potential $V$
is polynomial and  consider the cases of any  multi-cut support of the equilibrium measure. It
is shown that fluctuations become gaussian in the limit $n\to\infty$, if they are normalized by $\pi^{-2}\log n$.
\end{abstract}

%

\section{Introduction and main results}\label{s:1}

Consider $\beta$-ensemble of random matrices, whose joint eigenvalue distribution
is
\begin{align}\label{p(la)}
p_{n,\beta}(\lambda_1,...,\lambda_{n})=&Q_{n,\beta}^{-1}[V]\prod_{i=1}^n
e^{-n\beta V(\lambda_i)/2}\prod_{1\le i<j\le
n}|\lambda_i-\lambda_j|^\beta,
\end{align}
where $Q_{n,\beta}[V]$ is a normalizing factor
\begin{align*}&Q_{n,\beta}[V]=\int \prod_{i=1}^n
e^{-n\beta V(\lambda_i)/2}\prod_{1\le i<j\le
n}|\lambda_i-\lambda_j|^\beta d\bar\lambda.
\end{align*}
The function $V$ (called the potential) is a real valued H\"{o}lder function satisfying the condition
\begin{equation}\label{condV}
V(\lambda )\ge 2(1+\epsilon )\log(1+ |\lambda |).
\end{equation}
Below we denote
\begin{equation}\label{E}
    {E}\{(\dots)\}=\int(\dots)p_{n,\beta}(\lambda_1,...,\lambda_{n})d\bar\lambda.
\end{equation}
  This distribution can be considered
for any $\beta>0$, but the cases $\beta=1,2,4$ are especially important, since they
correspond  to real symmetric, hermitian, and symplectic
matrix models respectively.

It is known (see \cite{BPS:95,Jo:98}) that if $V'$ is a H\"{o}lder function, then
the empirical spectral distribution
\[ n^{-1}\sum_{j=1}^n \delta(\lambda-\lambda_i)\]
 converges weakly in probability defined by (\ref{p(la)}) to the function $\rho$ (equilibrium density) with a compact support
$\sigma$. The density $\rho$  maximizes the
functional, defined on the class $\mathcal{M}_1$ of positive unit measures on $\mathbb{R}$
\begin{equation}\label{E_V}
\mathcal{E}_V(\rho)=\max_{m\in\mathcal{M}_1}\bigg\{\int\log|\lambda-\mu|dm(\lambda) dm(\mu)
-\int V(\lambda)m(d\lambda)\bigg\}=\mathcal{E}[V].
\end{equation}
The support $\sigma$ and the density $\rho$ are uniquely defined by the conditions:
\begin{equation}\label{cond_rho}\begin{array}{l}\displaystyle
v(\lambda ):=2\int \log |\mu -\lambda |\rho (\mu )d\mu -V(\lambda )=\sup v(\lambda):=v^*,\quad\lambda\in\sigma,\\
v(\lambda )\le \sup v(\lambda),\quad \lambda\not\in\sigma,\hskip
2cm\sigma=\hbox{supp}\{\rho\}.
\end{array}\end{equation}

We are interested in the behavior of linear eigenvalue statistics, i.e.,
\begin{equation}\label{Linst}
\mathcal{N}_n[h]=\sum_{j=1}^nh(\lambda_j^{(n)}),
\end{equation}
In the case of smooth test function $h$ the behavior of $\mathcal{N}_n[h]$ now is very well understood for any $\beta>0$.
It was proven in \cite{Jo:98}  that for  one cut (i.e., $\sigma=[a,b]$) polynomial  potentials of generic behavior and
sufficiently smooth  $h$ (8 derivatives), if we
 consider  the
characteristic functional of $\mathcal{N}_n[h]$ in the form
\begin{align}\label{Phi_1} \Phi_{n,\beta}[x,h]=& {E}\Big\{e^{x(\mathcal{N}_n[h]
-E\{\mathcal{N}_n[h])\}}\Big\},
\end{align}
then
\begin{align*}
&  \lim_{n\to\infty}\Phi_{n,\beta}[h]=\exp\Big\{\frac{x^2}{2\beta}(\overline D_\sigma h,h)\Big\},
\end{align*}
where the "variance operator" $ \overline D_\sigma$ and the measure $\nu$ have the form
\begin{align*}& (\overline D_\sigma h,h)=
\int_{\sigma} \frac{h(\lambda)d\lambda}{\pi^2X^{1/2}_{\sigma}(\lambda)}\int_{\sigma}
\frac{h'(\mu)X^{1/2}_{\sigma}(\mu) d\mu}{\lambda-\mu}, \\
&X_\sigma(\lambda)=(b-\lambda)(\lambda-a).
\end{align*}
 The method of \cite{Jo:98} was improved  in \cite{KS:10}, where it was generalized to the case of non polynomial
 real analytic potentials $V$
 and the test functions with 4 derivatives, and then improved once more in \cite{S:13},
 where the case of non analytic $V$ was also studied.
 The case of multi-cut (i.e. $\sigma$ consisting of more than one interval) real analytic potentials was studied in \cite{S:14}, where it was shown that in this case fluctuations
 become non gaussian.

 But the method, used in the case of smooth $h$, does not work in the case of  $h$ which have  jumps.
 In particular, the method is not applicable to $h=1_{\Delta}$, $\Delta=[a,b]\subset\sigma$,
 which means that $\mathcal{N}_n[h]$ is a number
 of eigenvalues inside the interval $\Delta$. Moreover,  it is known that for gaussian unitary and gaussian orthogonal ensembles
 ($GUE$ and $GOE$)  the variance of the eigenvalue number  is proportional to $\log n$, while in the case of smooth test functions
 the variance is $O(1)$. Thus, it is hard to believe that the central limit theorem (CLT)
  for indicators can be obtained by  methods similar to
 that for smooth test functions.

 Till now there are only few results on the CLT for indicators. The case of GUE
 was studied a long time ago (see,e.g., \cite{Me:91}). In the paper \cite{Co-Leb:95} it was shown that the Gaussian
 fluctuations for GUE imply similar results for GOE and GSE (i.e., the cases when $V(\lambda)=\lambda^2/2$ and $\beta=1$ and
 $\beta=4$). Even for classical random matrix models, like the Wigner model with non gaussian entries,
 CLT for  functions with jumps was  proven  (see \cite{Pan:12}) only for the Hermitian case ($\beta=2$),  and only
 under the assumption that the  first four moments of the entries coincide with that of GUE.
 There are also a number of publications
 where CLT for the determinantal point processes are proven (see \cite{So:00} and references therein or \cite{Fo-Leb:13}).
  Similar results for
 some special kind of Pfaffian point processes  were obtained in \cite{K:13}.

 At the present paper we use the representation
 of the characteristic functional of $\mathcal{N}_n[h]$ in the form of the Fredholm determinant of some operator
 in order to prove CLT for the indicator test functions in the case of $\beta$- models
 with $\beta=1,2,4$. Unfortunately, since
  similar representations are not known for general $\beta$, the method does not work for $\beta\not=1,2,4$

Let us start form the case $\beta=2$.

Given potential $V$, introduce the weight function $ w_{n}(\lambda )=e^{-n V(\lambda )}$,
and consider polynomials orthogonal  on $\mathbb{R}$ with the weight $ w_{n}$
 i.e.,
\begin{equation}\label{ortP}
\int p_{l}^{(n)}(\lambda )p_{m}^{(n)}(\lambda )w_{n}(\lambda )d\lambda
=\delta _{l,m}.
\end{equation}%
It will be used below also that  $\{p_{l}^{(n)}\}_{l=0}^n$ satisfy the recursion relation
\begin{equation}\label{rec}
\lambda p_{l}^{(n)}(\lambda )=a_{l+1}p_{l+1}^{(n)}(\lambda )+ b_np_{l}^{(n)}(\lambda )+a_{l}p_{l}^{(n)}(\lambda ).
\end{equation}

Then consider the orthonormalized system
\begin{equation}\label{psi}
\psi _{l}(\lambda )=e^{-n V(\lambda )/2}p_{l}^{(n)}(\lambda
),\;\,l=0,...,
\end{equation}
and construct the function
\begin{equation}\label{K_2}
K_{n}(\lambda ,\mu )=\sum_{l=0}^{n-1}\psi _{l}(\lambda )\psi
_{l}(\mu ).
\end{equation}%
This function is known as a reproducing kernel of the system (\ref{psi}).
It is  known (see, e.g., \cite{Me:91}) that for any $x$ and any bounded integrable test functions $h$
the characteristic functional
 defined by (\ref{Phi_1}) for $\beta=2$ takes the form
\[\Phi_{n,2}[x,h]=e^{-xE\{\mathcal{N}_n[h]\}}\mathrm{det}\Big\{1+(e^{xh}-1)K_n\Big\},\]
where the operator  $(e^{xh}-1)K_n$ has the kernel
\[((e^{xh}-1)K_n)(\lambda,\mu):=(e^{xh(\lambda)}-1)K_n(\lambda,\mu)\]
In particular, if $h=1_\Delta$ and we set $x_n=x\pi/\log^{1/2} n$, then $\Phi_{n,2}[x_n,1_\Delta]$ takes the form
\begin{align}\label{hat_Phi}
\hat\Phi_{n,2}(x):=e^{-x_nE\{\mathcal{N}_n[1_\Delta]\}} E\{e^{x_n\mathcal{N}_n[1_\Delta]}\}
=e^{-x_nE\{\mathcal{N}_n[{1}_{\Delta}]\}}\mathrm{det}\Big\{1+(e^{x_n}-1)K_n[\Delta]\Big\},
\end{align}
where
\begin{align}\label{K[D]}
K_n[\Delta](\lambda,\mu):={1}_{\Delta}(\lambda)K_{n}(\lambda ,\mu ){1}_{\Delta}(\mu).
\end{align}
Representation (\ref{hat_Phi}) allows us to prove CLT for the indicator test function in the case $\beta=2$
(see, e.g., \cite{PS:11}):
\begin{theorem}\label{t:2}
Let the  matrix model be defined by (\ref{p(la)}) with $\beta=2$ and  real analytic potential $V(\lambda)>>\log|\lambda^2+1|$.
Let also $\Delta=[a,b]\subset\sigma^\circ$ (here and below $\sigma^\circ$ means the internal part of the
support $\sigma$ of the equilibrium measure) and
 $x_n=x\pi\log^{-1/2} n$. Then
\begin{align}\label{t2.0}
\lim_{n\to\infty}\log\hat\Phi_{n,2}(x)=x^2/2.
\end{align}
\end{theorem}
Although the result is not new, its proof is an important ingredient of the proofs of CLT for the cases $\beta=1,4$,
 hence the proof is given in the beginning of Section 2.

For $\beta=1,4$ the situation is more complicated. It was shown in \cite{Tr-Wi:98} that
the characteristic functionals $\hat\Phi_{n,1}(x)$ and
$\hat\Phi_{n,4}(x)$ can be expressed in terms of some matrix kernels (see (\ref{Kn1}) --
(\ref{D,M}) below). But the representation is less convenient than (\ref{ortP}) -- (\ref{hat_Phi}).
It makes difficult the problems, which for $\beta=2$ are just simple exercises.

We have
\begin{align}\label{hat_Phi_1,4}
\hat\Phi_{n,1}(x)
&=e^{-x_nE\{\mathcal{N}_n[{1}_{\Delta}]\}}\mathrm{det}^{1/2}\Big\{1+(e^{x_n}-1)K_{n,1}[\Delta]\Big\},\\
\hat\Phi_{n,4}(x)
&=e^{-x_nE\{\mathcal{N}_{n/2}[{1}_{\Delta}]\}}\mathrm{det}^{1/2}\Big\{1+(e^{x_n}-1)K_{n,4}[\Delta]\Big\},
\notag\end{align}
where similarly to the case $\beta=2$ the operators $K_{n,1}[\Delta]$ and
$K_{n,4}[\Delta]$ are the projection on the interval $\Delta$ of some matrix operators $K_{n,1}$, $K_{n,4}$:
\[K_{n,1}[\Delta](\lambda,\mu)=1_\Delta(\lambda)K_{n,1}(\lambda,\mu)1_\Delta(\mu),\quad
K_{n,4}[\Delta](\lambda,\mu)=1_\Delta(\lambda)K_{n,4}(\lambda,\mu)1_\Delta(\mu).\]

The matrix operators $K_{n,1}$ and $K_{n,4}$ have the form
\begin{align}\label{Kn1}
     K_{n,1} &:=
     \begin{pmatrix}
          \mathcal{S}_{n,1} & \mathcal{D}_{n,1}\\
          \mathcal{I}_{n,1} -\epsilon & \mathcal{S}^T_{n,1}
     \end{pmatrix},\quad\beta=1,\,\, n-\mathrm{even},\\
     \label{Kn4}
     K_{n,4} &:=\frac{1}{2}
     \begin{pmatrix}
          \mathcal{S}_{n,4} & \mathcal{D}_{n,4}\\
        \mathcal{I}_{n,4}  & \mathcal{S}^T_{n,4}
     \end{pmatrix},\quad \beta=4,
\end{align}
where the entries are integral operators in $L_2[\mathbb{R}]$ with the kernels
\begin{align}\label{Sn1}
     \mathcal{S}_{n,1}(\lambda,\mu) &= -\sum_{j,k=0}^{n-1} \psi_j(\lambda)
     (M_n^{(n)})_{jk}^{-1} (\epsilon \psi_k)(\mu),\quad\mathcal{S}^T_{n,1}(\lambda,\mu)=\mathcal{S}_{n,1}(\lambda,\mu),\\
    \mathcal{D}_{n,1}(\lambda,\mu)&=-\frac{\partial}{\partial \mu} \mathcal{S}_{n,1}(\lambda,\mu),\quad
\mathcal{I}_{n,1}(\lambda,\mu)=(\epsilon \mathcal{S}_{n,1}) (\lambda,\mu),\notag\\
\label{Sn4}
    \mathcal{S}_{n/2,4}(\lambda,\mu) &= -\sum_{j,k=0}^{n-1} \psi_j'(\lambda)
     (D_n^{(n)})_{jk}^{-1}  \psi_k^{(n)}(\mu),\quad \mathcal{S}^T_{n/2,4}(\lambda,\mu)=\mathcal{S}_{n/2,4}(\lambda,\mu),\\
\mathcal{D}_{n,4}(\lambda,\mu)&=-\frac{\partial}{\partial \mu} \mathcal{S}_{n,4}(\lambda,\mu),\quad
\mathcal{I}_{n,4}(\lambda,\mu)=(\epsilon \mathcal{S}_{n,4}) (\lambda,\mu),\notag\\
 \epsilon(\lambda-\mu)&=\frac{1}{2} \hbox{sgn} (\lambda-\mu).
\end{align}
Here the function $\{\psi_j\}_{j=0}^n$ are defined by (\ref{psi}), $\hbox{sgn}$ denotes the standard signum function, and
$D^{(n)}_n$ and $M^{(n)}_n$  in (\ref{Sn1}) and (\ref{Sn4}) are the left top corner $n\times n$ blocks of the
semi-infinite matrices  that correspond to the differentiation operator and to some integration operator respectively.
\begin{align}\label{D,M}
     D^{(n)}_\infty &:= (\psi_j',\psi_k)_{j,k\ge 0},
     \quad D^{(n)}_n=\{D^{(n)}_{jk}\}_{j,k=0}^{n-1},\\
     M_\infty^{(n)} &:= \left (\epsilon \psi_j,\psi_k\right)_{j,k\ge 0},\
     \quad M^{(n)}_n=\{M^{(n)}_{jk}\}_{j,k=0}^{n-1}.
\notag\end{align}
\begin{remark}
From the structure of the kernels it is easy to see  the cases $\beta=1,4$ the characteristic functional can be written
in the form
\begin{align*}
\hat\Phi_{n,4}(x)
=\mathrm{det}^{1/2}\Big\{J+(e^{x_n}-1)\hat A_{n,1}[\Delta]\Big\}e^{-x_nE\{\mathcal{N}_n[{1}_{\Delta_a}]\}},\\
\hat\Phi_{n,4}(x)
=\mathrm{det}^{1/2}\Big\{J+(e^{x_n}-1)\hat A_{n,4}[\Delta]\Big\}e^{-x_nE\{\mathcal{N}_n[{1}_{\Delta_a}]\}},\\
\end{align*}
where $\hat A_{n,1}=\mathcal{S}_{n,1}J$, $\hat A_{n,4}=\mathcal{S}_{n,4}J$ are skew symmetric matrices
\begin{align*}
 (A_{n,1}[\Delta])^*=-A_{n,1}[\Delta], \quad  (A_{n,4}[\Delta])^*=-A_{n,4}[\Delta],\quad
 J=\left(\begin{array}{rr}0&I\\-I&0\end{array}\right).
\end{align*}
\end{remark}

The main problem of studying of $\hat\Phi_{n,1}(x)$ and
$\hat\Phi_{n,4}(x)$ is that  the corresponding operators $K_{n,1}$ and $K_{n,4}$
(differently from the case $\beta=2$) are not self adjoint, thus even if we know  the location of eigenvalues of $K_{n,1}$ and $K_{n,4}$
we cannot say something about the location of eigenvalues of $K_{n,1}[\Delta]$ and $K_{n,4}[\Delta]$.

The idea is to prove that the eigenvalue problems for $K_{n,1}[\Delta]$ and $K_{n,4}[\Delta]$ can be reduced
to the eigenvalue problem for $K_{n}[\Delta]$ with some finite rank perturbation. For this aim we
use the result of  \cite{Wi:99}, where it was observed
that if  $V$ is a rational function,  in particular, a polynomial of degree $2m$,
then the kernels $\mathcal{S}_{n,1},\mathcal{S}_{n,4}$ can be written as
\begin{align}\label{Sn.1}
     \mathcal{S}_{n,1}(\lambda,\mu) &= K_{n}(\lambda,\mu)+n\sum_{j,k=-(2m-1)}^{2m-1} F^{(1)}_{jk}\psi_{n+j}(\lambda)
     \epsilon\psi_{n+k}(\mu),\\
     \mathcal{S}_{n/2,4}(\lambda,\mu)&= K_{n} (\lambda,\mu) +n\sum_{j,k=-(2m-1)}^{2m-1} F^{(4)}_{jk}\psi_{n+j}(\lambda)
     \epsilon\psi_{n+k}(\mu),
\notag\end{align}
where  $F_{jk}^{(1)}$, $F_{jk}^{(4)}$ can be expressed in terms of the matrix $T_n^{-1}$,
where $T_n$ is the
$(2m-1)\times (2m-1)$ block in the bottom right corner of $D_n^{(n)}M_n^{(n)} $, i.e.,
\begin{equation}\label{T}
     (T_n)_{jk}:= (D_n^{(n)}M_n^{(n)})_{n-2m+j, n-2m+k},\quad\quad 1\le j, k\le 2m-1.
\end{equation}
The representation was used  before to study local regimes for real symmetric and symplectic matrix models.
The main technical obstacle there was the problem to prove
that $(T_n^{-1})_{jk}$  are bounded uniformly in $n$. The problem was solved initially for the case
of monomial $V(\lambda)=\lambda^{2m}$ in \cite{De-G:07}, then for general one-cut real analytic $V$ in \cite{S:09}
and finally for the general multi cut potential in \cite{S:11}, where it was shown that for generic real analytic potential $V$
\begin{equation}\label{F}
    |F^{(1)}_{jk} |\le C,\quad |F^{(4)}_{jk} |\le C.
\end{equation}

 To formulate the main results, let us state our  conditions.

\medskip \noindent \textit{ C1. }\textit{$V$ is a polynomial of degree $2m$ with a positive leading
coefficient, and the support
of its equilibrium measure is}
\begin{equation}\label{sigma}
    \sigma=\bigcup_{\alpha=1}^q\sigma_\alpha,\quad \sigma_\alpha=[E_{2\alpha-1},E_{2\alpha}]
\end{equation}
\medskip \noindent \textit{ C2. }\textit{ The equilibrium density $\rho$ can be represented in the form
\begin{equation}\label{rho}
    \rho(\lambda)=\frac{1}{2\pi}P(\lambda)\Im X^{1/2}(\lambda+i0),\quad
    \inf_{\lambda\in\sigma}|P(\lambda)|>0,
\end{equation}
where
\begin{equation}\label{X}
    X(z)= \prod_{\alpha=1}^{2q}(z-E_\alpha),
    \end{equation}
and we choose a branch of $X^{1/2}(z)$ such that $X^{1/2}(z)\sim z^q$, as $z\to+\infty$.
Moreover, the function $v$ defined by (\ref{cond_rho})
attains its maximum only if $\lambda $ belongs to  $\sigma $. }

\begin{remark} It is known (see, e.g., \cite[Theorem 11.2.4]{PS:11}) that for any analytic $V$
the equilibrium density $\rho$ always
 has the form (\ref{rho}) -- (\ref{X}).
 The function $P$ in
(\ref{rho}) is  analytic and  can be represented in the form
\begin{equation*}
    P(z)=\frac{1}{2\pi i}\oint_\mathcal{L}\frac{V'(z)-V'(\zeta)}{(z-\zeta) X^{1/2}(\zeta)}d\zeta.
    \end{equation*}
Hence  condition C2 means that $\rho$ has no zeros in the internal points of $\sigma$ and behaves like
square root near the edge points. This behavior of $\rho$ is usually called generic.
\end{remark}

\begin{theorem}\label{t:3} Consider the matrix model (\ref{p(la)}) with $\beta=1$ and even $n$
and $V$ satisfying conditions C1,C2. Let the interval $\Delta=[a,b]\subset\sigma^\circ$, and
let the characteristic functional $\hat\Phi_{n,1}(x)$ be defined by (\ref{hat_Phi}) for $\beta=1$ with $x_n=x\pi\log^{-1/2}n$.
 Then
 \[\lim_{n\to\infty}\log\hat\Phi_{n,1}(x)=x^2.\]
\end{theorem}
 \begin{theorem}\label{t:4}Consider the matrix model (\ref{p(la)}) with $\beta=4$
and $V$ satisfying conditions C1,C2. Let the interval $\Delta=[a,b]\subset\sigma^\circ$ and
 let characteristic functional $\hat\Phi_{n,4}(x)$ be defined by (\ref{hat_Phi}) for $\beta=4$ with $x_n=x\pi\log^{-1/2}n$.
 Then
 \[\lim_{n\to\infty}\log\hat\Phi_{n,4}(x)=x^2/4.\]
\end{theorem}

\section{Proofs}

\textit{Proof of Theorem \ref{t:2}} Set $F(x_n):=\log\hat\Phi_{n,2}(x)$ and consider the Taylor expansion
of $F(x_n)$ with respect to $x_n$ up to the second order
\begin{align}\label{t2.1}
F(x_n)=
&\frac{x_n^2}{2}\mathrm{Tr}\,\,K_n[\Delta](1-K_n[\Delta])+\frac{x_n^3}{6}\mathrm{Tr}\,\,K_n[\Delta](1-K_n[\Delta])
\tilde R(K_n[\Delta]),\\
&C_1\le \tilde R(t)\le C_2, \quad t\in[0,1].
\notag\end{align}
\begin{lemma}\label{l:1}
\begin{align}\label{p1.1}
\mathrm{Tr}\,\,K_n[\Delta](1-K_n[\Delta])=\int_{\Delta}d\lambda\int_{\bar\Delta}K_n^2(\lambda,\mu )d\mu=\pi^{-2}\log n(1+o(1)).
\end{align}
\end{lemma}
The lemma implies that the first term in the r.h.s. of (\ref{t2.1}) tends to $x^2/2$, while the second one is bounded
by $cx_n^3\log n=o(1)$, since
\[C_1\mathrm{Tr}\,\,K_n[\Delta](1-K_n[\Delta])\le\mathrm{Tr}\,\,K_n[\Delta](1-K_n[\Delta])\tilde R(K_n[\Delta])
\le C_2\mathrm{Tr}\,\,K_n[\Delta](1-K_n[\Delta]).\]
Hence, we get the assertion of Theorem \ref{t:2}. Thus, we are left to prove Lemma \ref{l:1}

\textit{Proof of Lemma \ref{l:1}.}
Take $d_n=\log^{1/3}n$ and write
\begin{align}\label{p1.2}
\mathrm{Tr\,}K_n[\Delta](1-K_n[\Delta])=
&\int_{\Delta}d\lambda\int_{\bar\Delta}d\mu K_n^2(\lambda,\mu)=\Big(\int_{a}^{a+d_n}d\lambda+\int_{a+d_n}^{b-d_n}d\lambda+
\int_{b-d_n}^bd\lambda\Big)\\
\times&\Big(\int_{a-d_n}^{a}d\mu+\int_b^{b+d_n}d\mu+\int_{-\infty}^{a-d_n}d\mu+\int_{b+d_n}^\infty d\mu\Big)
K_n^2(\lambda,\mu).
\notag\end{align}
The Christoffel-Darboux formula implies
\[\int K_n^2(\lambda,\mu)(\lambda-\mu)^2d\lambda d\mu=a_n\int(\psi_n(\lambda)\psi_{n-1}(\mu)-
\psi_n(\mu)\psi_{n-1}(\lambda))^2d\lambda d\mu= 2a_n\le C,      \]
where $a_n$ is the recursion coefficient  of (\ref{rec}), and we have used  the result of \cite{PS:97}
 (see also \cite{PS:11}, Chapter, Lemma) on the uniform boundedness of $a_{n}$, as $n\to\infty$.
Then
\[\int_{d_n\le|\lambda-\mu|}K_n^2(\lambda,\mu)d\lambda d\mu\le Cd_n^{-2}=O(\log^{2/3} n),\]
which implies that
\begin{align}\label{p1.3}
\mathrm{Tr\,}K_n[\Delta](1&-K_n[\Delta])=
\int_{a}^{a+d_n}d\lambda\int_{a-d_n}^{a}d\mu K_n^2(\lambda,\mu)\\&+\int_{b-d_n}^bd\lambda\int_b^{b+d_n}d\mu
K_n^2(\lambda,\mu)+O(\log^{2/3} n)=I_a+I_b+O(\log^{2/3} n).
\notag\end{align}
To find $I_a$, we apply the results of \cite{DKMVZ:99},
according to which for $\lambda,\mu$ from the bulk of the spectrum the reproducing kernel has the form
\begin{align}\label{p2.3}
K_n(\lambda,\mu)=&h(\lambda,\mu)\frac{\sin n\pi (\phi(\lambda)-\phi(\mu))}{\pi(\lambda-\mu)}(1+O(n^{-1}))\\&+
\sum_{\pm} r_{\pm,\pm}(\lambda,\mu)e^{i\pi n(\pm\phi(\lambda)\pm\phi(\mu))},
\notag\end{align}
where $h$ and $\phi$ for $(\lambda,\mu)$ in the bulk of the spectrum are smooth, positive, bounded from both sides functions,
the remainder functions $r_{+,+}$, $r_{-,-}$, $r_{+,-}$, $r_{-,+}$ have uniformly bounded  derivatives in both variables,
and $\sum_{\pm}$ means the summation  with respect to all combinations of signs
in the exponents.
Moreover,
\begin{align*}
&\phi'(\lambda)>c_0,\quad\mathrm{if}\quad  |\lambda-E_k|\ge\tilde\varepsilon,\quad k=1,\dots,2q,\\
&h(\lambda,\lambda)=1.
\end{align*}
It is easy to see that the remainder terms in the r.h.s. of (\ref{p2.3}) after integration
 in the limits, written in the r.h.s. of (\ref{p1.3}), give us at most $O(d^2_n)$. Hence, we need only to find the contribution
of the first term of (\ref{p2.3}).
Performing the change of variables $\lambda=a+x/(n\phi'(a))$,
$\mu=a-y/(n\phi'(a))$, we get
\begin{align*}
I_a=&\int_0^{nd_n}dx\int_0^{nd_n}dy (1+o(1))\frac{\sin^2(\pi(x+y)(1+o(1))}{\pi^2(x+y)^2}dxdy+O(d^2_n)\\
=&\int_0^{nd_n}dx\int_0^{nd_n}dy \frac{\sin^2(\pi(x+y))}{\pi^2(x+y)^2}dxdy+o(\log{nd_n})
+O(d^2_n)\\
=&\Big(\int_0^1dx+\int_1^{nd_n} dx\Big)\Big(\int_0^1dy+\int_1^{nd_n} dy\Big)\frac{\sin^2 (\pi(x+y))}{\pi^2(x+y)^2}dy
+o(\log{nd_n})+O(d^2_n)
\\=&\int_1^{nd_n} dx\int_1^{nd_n} dy\frac{1-\cos 2\pi(x+y)}{2\pi^2(x+y)^2}+O(1)+o(\log{nd_n})+O(d^2_n)\\
=&\int_1^{nd_n} dx\int_1^\infty dy\frac{1}{2\pi^2(x+y)^2}+O(1)+o(\log{nd_n}))=\frac{1}{2\pi^{2}}\log n(1+o(1)).
\end{align*}
Similarly
\[I_b=\frac{1}{2\pi^{2}}\log n(1+o(1)).\]
Then in view of (\ref{p1.3}) we obtain (\ref{p1.1}).

$\square$

 \textit{Proof of Theorem \ref{t:3}}

Let us consider the eigenvalue problem for $\hat K_{n,1}[\Delta]$:
\begin{align}\label{ep.1}
&\left\{\begin{array}{l}\mathcal{S}_{n,1}f_{\Delta }+\mathcal{D}_{n,1}g_{\Delta }=Ef_{\Delta },\\
\mathcal{I}_{n,1}f_{\Delta }-\epsilon f_{\Delta }+\mathcal{S}_{n,1}^Tg_{\Delta }=Ef_{\Delta }.\end{array}\right.\end{align}
Here and below
\[f_{\Delta }=1_{\Delta }f,\quad g_{\Delta }=1_{\Delta }g.\]
Observe, that since all the functions in the first line of (\ref{ep.1}) are analytic, the equation is
valid also outside of $\Delta$.
Apply the operator $\epsilon$ to both sides of the equation. We get
\begin{align}\label{ep.2}&
\left\{\begin{array}{l}\mathcal{I}_{n,1}f_{\Delta }+\mathcal{S}_{n,1}^Tg_{\Delta }=E\epsilon f_{\Delta }
+E\epsilon f_{\bar\Delta},\\
\mathcal{I}_{n,1}f_{\Delta }-\epsilon f_{\Delta }+\mathcal{S}_{n,1}^Tg_{\Delta }=Eg_{\Delta },\end{array}\right.\\
&\Rightarrow
Eg_{\Delta }=(E-1)(\epsilon f_{\Delta })+\mathbf{1}_{\Delta}E\epsilon f_{\bar\Delta },
\notag\end{align}
where we use that integration by parts gives us that $\epsilon\mathcal{D}_{n,1}=\mathcal{S}^T_{n,1}$, and denote
\[f_{\bar\Delta}=f-f_{\Delta}\]
with $\bar\Delta$ being a complement of $\Delta$. Observe that
\begin{align}\label{ep.3}
\epsilon f_{\bar\Delta}(\lambda)=\frac{1}{2}\int_{-\infty}^af(t)dt-\frac{1}{2}\int_b^{\infty}f(t)dt
=:(f,\Psi_\Delta)=\mathrm{const},\quad
\lambda\in\Delta.
\end{align}
Multiply the first line of (\ref{ep.2}) by $E$ and use the above equation for $Eg_{\Delta }$. Integration by parts
gives us
\begin{align}\notag
&\mathcal{D}_{n,1}(\epsilon f_{\Delta })(\lambda)=\mathcal{S}_{n,1}f_{\Delta }(\lambda),\\
& \mathcal{D}_{n,1}(\mathbf{1}_{\Delta}\epsilon f_{\bar\Delta })(\lambda)
=(\mathcal{S}_{n,1}(\lambda,a)-\mathcal{S}_{n,1}(\lambda,b))(f,\Psi_\Delta)=:Pf,
\label{P}\end{align}
where by the above definition, $P$ is a rank one operator in $L_2[\Delta]$. We obtain
\begin{align*}
(2E-1)\mathcal{S}_{n,1}[\Delta]f_{\Delta }-E^2f_{\Delta }+EPf=0.
\end{align*}
Hence the solutions $\{E_k\}$ of (\ref{ep.1}) are solutions of the equation
\begin{align*}
\mathcal{P}(E):=\mathrm\det\Big\{E^2-(2E-1)\mathcal{S}_{n,1}[\Delta]+EP\Big\}=0.
\end{align*}

It is evident that $\mathcal{P}(E)$ is a polynomial of $2n$th degree, and $E_k$ are the roots of $\mathcal{P}(E)$.
We are interested in
\begin{align*}
\prod_{k=1}^{2n}(1+\delta_n E_k)=\delta_n^{2n}\prod_{k=1}^{2n}(\delta_n^{-1}+E_k)=\delta_n^{2n}\mathcal{P}(-\delta_n^{-1}),
\end{align*}
where $\delta_n:=e^{x_n}-1$. Thus we obtain
\begin{align}\label{ep.4}
\log\hat\Phi_{n,1}(x)
=&-x_nE\{\mathcal{N}_n[\mathbf{1}_{\Delta }]\}+\frac{1}{2}\log\mathrm{det}\Big\{1+
(2\delta_n+\delta_n^2)\mathcal{S}_{n,1}[\Delta]+\delta_nP\Big\}.
\end{align}
Now we use (\ref{Sn.1}). Substituting the representation in (\ref{ep.4}) we get
\begin{align}\label{ep.5}
\log\hat\Phi_{n,1}(x)
=&-x_nE\{\mathcal{N}_n[\mathbf{1}_{\Delta }]\}+\frac{1}{2}\log\mathrm{det}\Big\{1+(2\delta_n+\delta_n^2)K_n(\Delta)\Big\}
\\&+\frac{1}{2}\log\mathrm{det}\Big\{(1+R(\delta_nP_1+\tilde\delta_nn\sum_{k,j=-(2m-1)}^{2m-1} F^{(1)}_{kj}Q_{kj}\Big\},
\notag\end{align}
where $\{Q_{kj}\}$ are rank one operators with the kernels $Q_{kj}(\lambda,\mu)=\psi_{n+k}(\lambda)\epsilon\psi_{n+j}(\mu)$,
\[
R=(1-\tilde\delta_nK_n[\Delta])^{-1},\quad \tilde\delta_n=(e^{2x_n}-1)=2\delta_n+\delta_n^2.
\]
According to the standard linear algebra  argument
\[
\mathrm{det}(1+\tilde\delta_n\sum a_i\otimes b_i)=\mathrm{det}\big\{\delta_{ij}+\tilde\delta_n(a_i,b_j)\big\}.
\]
Taking into account the formula and the structure of the remainder in (\ref{ep.5}), we conclude that in order to prove that
the last term in (\ref{ep.5}) is small, it suffices to prove that
\begin{align}\label{ep.6}
&|n(R\psi_{n-j},\epsilon\psi_{n+k})|\le C\delta_n,\\
& |(R\mathbf{1}_\Delta S_n(x,a),\Psi_\Delta)|\le C\delta_n,\quad|(R\mathbf{1}_\Delta S_n(x,b),\Psi_\Delta)|\le C\delta_n.
\notag\end{align}
The last two inequalities are trivial, since
\[
\mathrm{supp}\,\Psi_\Delta=\bar\Delta,\quad
\mathrm{supp}\,R{1}_\Delta S_n(x,a)=\mathrm{supp}\,R{1}_\Delta S_n(x,b)=\Delta.
\]
For the proof of the first  inequality of (\ref{ep.6}) we need the following lemma.
\begin{lemma}\label{l:2} Set
\begin{align}\label{v_n}
&v_n(\lambda):=1_{\bar\Delta}(\lambda)\int_\Delta d\mu K_n(\lambda,\mu).
\end{align}
Then for any $\lambda\in\bar\Delta$
\begin{align}\label{p.2}
&|v_n(\lambda)|\le \frac{C}{1+n|\lambda-b|}+
 \frac{C}{1+n|\lambda-a|}.
\end{align}
\end{lemma}
The proof of the lemma is given after the proof of Theorem \ref{t:3}. Now we continue the proof of (\ref{ep.6}).
The first bound of (\ref{ep.6}) is a corollary of three estimates
\begin{align}\label{ep.7}
&|n(1_\Delta \psi_{n+k},\epsilon\psi_{n+j})|\le C\delta_n,\quad
|n(\psi_{n+k},K_n[\Delta]\epsilon\psi_{n+j})|\le C\delta_n,\\&
\|n(K_n[\Delta]-K_n^2[\Delta])\epsilon\psi_{n+j})\|\le C.
\notag\end{align}
Indeed, the third bound of (\ref{ep.7}) yields for $m\ge 2$
\begin{align*}
&\|n(K_n^m[\Delta]-K_n[\Delta])\epsilon\psi_{n+j})\|\le
\sum_{l=0}^{m-1} \|K^l_n[\Delta]\,n(K_n^2[\Delta]-K_n[\Delta])\epsilon\psi_{n+j}\|\\
&\le m\|n(K_n^2[\Delta]-K_n[\Delta])\epsilon\psi_{n+j})\|\le mC.
\end{align*}
Here we used also that $\|K_n[\Delta]\|\le 1$. Thus,
\begin{align*}
&\Big\|\sum_{m=2}^\infty\tilde\delta_n^mn(K_n^m[\Delta]-K_n[\Delta])\epsilon\psi_{n+j})\Big\|\le
 \sum_{m=2} mC\tilde\delta_n^m\le C\tilde\delta_n^2,\\
 &\Rightarrow\Big\|\sum_{m=2}^\infty n\tilde\delta_n^mK_n^m[\Delta]\epsilon\psi_{n+j}-
\frac{n(2\tilde\delta_n^2-\tilde\delta_n^3)}{(1-\tilde\delta_n)^2}K_n[\Delta]\epsilon\psi_{n+j}\Big\|\le C\tilde\delta_n^2.
\end{align*}
Combining this inequality with the first two bounds of (\ref{ep.7}) we obtain the first bound of (\ref{ep.6}).

To prove (\ref{ep.7}), we use the result of \cite[Lemma 2]{S:11}, according to which
\begin{align*}
\epsilon\psi_{n+j}=n^{-1/2}c_{n+j}+O(n^{-1}),
\end{align*}
where $c_{n+j}$ is some constant, bounded uniformly in $n$.  Using this fact, we conclude that to prove (\ref{ep.7})
it suffices to prove that
\begin{align}\label{ep.8}
&|n^{1/2}(1_\Delta \psi_{n+k},1_\Delta)|\le C\delta_n,\quad
|n^{1/2}(K_n[\Delta]\psi_{n+k},1_\Delta)|\le C\delta_n,\\&
\|n^{1/2}(K_n[\Delta]-K_n^2[\Delta])1_\Delta)\|\le C.
\notag\end{align}
 Lemma \ref{l:2} yields
\begin{align*}
&\|(K_n[\Delta]-K_n^2[\Delta])1_\Delta\|=\Big\| \int_{\bar\Delta} d\nu \int_\Delta d\mu K_n(\lambda,\nu)K_n(\mu,\nu)\Big\|
\\&=\| K_nv_{n}\|\le\|v_{n}\|\le C_1 n^{-1/2},
\end{align*}
hence we obtain the last inequality of (\ref{ep.8}). The first inequality of (\ref{ep.8}) is a simple corollary of the
result of \cite[Theorem 1.1]{DKMVZ:99}, according to which
\[\psi_{n+k}(\lambda)=R_k(\lambda)\cos (n\pi \phi(\lambda)+m_k(\lambda))\;(1+O(n^{-1}))\]
where $R_k$ and $m_k$ are smooth functions. Using this result, we can integrate by parts and obtain the first
inequality of (\ref{ep.8}) (even with $Cn^{-1/2}$ in the r.h.s. instead of $C\delta_n$).
In addition, since
\[
K_n\psi_{n+k}=1_{k<0}\psi_{n+k},
\]
we have
\[
K_n[\Delta]\psi_{n+k}=1_{k<0}1_{\Delta}\psi_{n+k}-K_n1_{\bar\Delta}\psi_{n+k}.
\]
The bound of the first term is given by the first inequality of (\ref{ep.8}). For the second term write
\[
|(K_n1_{\bar\Delta}\psi_{n+k},1_\Delta)|=|(1_{\bar\Delta}\psi_{n+k},K_n1_\Delta)|
=|(1_{\bar\Delta}\psi_{n+k}, v_n)|\le Cn^{-1/2}.
\]
Hence, we complete the proof of the second inequality of (\ref{ep.8}).

It was explained above that (\ref{ep.8}) imply (\ref{ep.7}), which combined with (\ref{ep.5}) yields
\begin{align*}
&\log\hat\Phi_{n,1}(x)=
-x_nE\{\mathcal{N}_n[\mathbf{1}_{\Delta }]\}+\frac{1}{2}\log\mathrm{det}\Big\{1+(e^{2x_n}-1)K_n[\Delta]\Big\}+O(\delta_n).
\end{align*}
Then similarly to the case $\beta=2$ we have
\begin{align}\label{t3.f}
\frac{1}{2}\mathrm{Tr}\,\,\log(1+(e^{2x_n}-1)K_n[\Delta])=
&x_nE\{\mathcal{N}_n[\mathbf{1}_{\Delta }]\}+x_n^2\mathrm{Tr}\,\,K_n[\Delta](1-K_n[\Delta])\\&+
\frac{(2x_n)^3}{12}\mathrm{Tr}\,\,K_n[\Delta](1-K_n[\Delta])\tilde R(K_n[\Delta]).
\notag\end{align}
By Lemma \ref{l:1},
\begin{align*}
\mathrm{Tr}\,\,K_n[\Delta](1-K_n[\Delta])=\pi^{-2}\log n\,(1+o(1)),
\end{align*}
hence the limit of the second term of (\ref{t3.f}) is $x^2$ and the last term is  $O(\log^{-1/2}n)$.

$\square$

\bigskip

\textit{Proof of Lemma \ref{l:2}}
The proof is based on the representation (\ref{p2.3}). Integrating by parts, it is easy to see that
the contribution of the remainder terms (written in $\sum_{\pm}$) is at most $O(n^{-1})$. Hence we need to
consider only the contribution of the first term in the r.h.s. of (\ref{p2.3}). Take $\lambda<a$ and
consider the change of variables $x=\phi(a)-\phi(\lambda)$, $y=\phi(\mu)-\phi(a)$. Let $\varphi$ be the inverse function
of $\phi(x)-\phi(a)$ and $a-\lambda=\Delta_\lambda\ge 0$.
Then the main part of our integral takes the form
\begin{align}\label{p2.1}
F(\lambda):=&\int_0^{d} dy \varphi'(y)\tilde h_\lambda(y)\frac{\sin n\pi(x+y)}{\varphi(y)+\Delta_\lambda}\\=&
 \varphi'(0)\tilde h_\lambda(0)\int_0^{d} dy\frac{\sin n\pi(x+y)}{\varphi(y)+\Delta_\lambda}+O(n^{-1})\notag\\=&
 C\int_0^{nd} dy'\frac{\sin \pi(nx+y')}{n\varphi(y'/n)+n\Delta_\lambda}+O(n^{-1}),
\notag\end{align}
where
\[
d=\phi(b)-\phi(a),\quad \tilde h_\lambda(y)=h(\lambda,\varphi(y)),
\]
and we have used the fact that the function
\[
\frac{\varphi'(y)\tilde h_\lambda(y)-\varphi'(0)\tilde h_\lambda(0)}{y+\Delta_\lambda}
\]
has a bounded derivative, hence integration by parts with $\sin n\pi(x+y)$ gives us $O(n^{-1})$
\begin{align*}
F(\lambda):=&C\int_0^{1-\{nx\}} dy \frac{\sin \pi(\{nx\}+y)}{n\varphi(y'/n)+n\Delta_\lambda}
-C\sum_{k=1}^{nd}\Big(\int_{-\{nx\}}^{1-\{nx\}}dy
\frac{\sin \pi(\{x\}+y)}{n\varphi((y'+k)/n)+n\Delta_\lambda}\\&-
\int_{-\{nx\}}^{1-\{nx\}}\frac{\sin \pi(\{nx\}+y')}{n\varphi((y'+k+1)/n)+n\Delta_\lambda}dy'\Big).
\end{align*}
Observe that the series above is of alternating sign, and modules of the terms decay, as $k$ grows (recall that $\varphi(y)$ is an increasing
function of $y$). Thus,
\begin{align*}
F(\lambda)\ge C&\int_0^{1-\{nx\}} dy' \frac{\sin \pi(\{nx\}+y')}{n\varphi(y'/n)+n\Delta_\lambda}-C\int_{-\{nx\}}^{1-\{nx\}}dy
\frac{\sin \pi(\{nx\}+y)}{n\varphi((y'+1)/n)+n\Delta_\lambda},\\
F(x)\le &C\int_0^{1-\{nx\}} dy\frac{\sin \pi(\{nx\}+y)}{(x+y)}-C\int_{-\{nx\}}^{1-\{nx\}}dy'
\frac{\sin \pi(\{nx\}+y')}{n\varphi((y'+1)/n)+n\Delta_\lambda}\\&+C\int_{-\{nx\}}^{1-\{nx\}}dy'
\frac{\sin \pi(\{nx\}+y')}{n\varphi((y'+2)/n)+n\Delta_\lambda}.
\end{align*}
These bounds combined with (\ref{p2.1}) prove (\ref{p.2}) for $\lambda<a$. For $\lambda>b$ the proof is the same.

$\square$

\bigskip

\textit{Proof of Theorem \ref{t:4}} The proof is very similar to that of Theorem \ref{t:3}, hence we present it very briefly.
We   consider the eigenvalue problem for $\hat K_{n,4}[\Delta]$ ($\beta=1)$.
\begin{align}\label{ep.1'}
&\left\{\begin{array}{l}\mathcal{S}_{n,4}f_{\Delta }+\mathcal{D}_{n,4}g_{\Delta }=Ef_{\Delta },\\
\mathcal{I}_{n,4}f_{\Delta }+\mathcal{S}_{n,4}^Tg_{\Delta }=Ef_{\Delta }.
\end{array}\right.\end{align}
Apply the operator $\epsilon$ to both sides of the first equation and then subtract the second line from the first. We get
\begin{align}\label{ep.2'}
Eg_{\Delta }=E(\epsilon f_{\Delta })+\mathbf{1}_{\Delta}E\epsilon f_{\bar\Delta }.
\notag\end{align}
Substituting the relation in the first line of (\ref{ep.1'}), we obtain
\begin{align*} 2\mathcal{S}_{n,4}[\Delta]f_{\Delta }-Ef_{\Delta }+Pf=0,
\end{align*}
where (cf (\ref{P}))
\begin{align*}
Pf:=(\mathcal{S}_{n,4}(\lambda,a)-\mathcal{S}_{n,4}(\lambda,b))(f,\Psi_\Delta)
\end{align*}
is a rank one operator. Taking into account (\ref{Sn4}) and (\ref{hat_Phi_1,4}), we have now (cf (\ref{ep.4}))
\begin{align*}
\log\hat\Phi_{n,4}(x)
=&-x_nE\{\mathcal{N}_n[\mathbf{1}_{\Delta }]\}
+\frac{1}{2}\log\mathrm{det}\Big\{1+\delta_n\mathcal{S}_{n,4}\Delta]+\frac{\delta_n}{2}P\Big\}.
\end{align*}
Applying (\ref{Sn.1}) and repeating the argument used in the proof of Theorem \ref{t:3}, we obtain the assertion of
Theorem \ref{t:4}.

$\square$


\begin{thebibliography}{99}

\begin{small}


\bibitem{BPS:95} Boutet de Monvel, A., Pastur L., Shcherbina M.: On the
statistical mechanics approach in the random matrix theory.
Integrated density of states.  J. Stat. Phys. \textbf{79},
585-611 (1995)

\bibitem{Pan:12}
 Bao, Z., Pan, G.,  Zhou, W.: Central limit theorem for partial linear eigenvalue statistics of Wigner matrices
 J.  Stat. Phys.  \textbf{ 150}(1),  88-129 (2013)

 \bibitem{Co-Leb:95} O. Costin and J.L. Lebowitz, Gaussian fluctuations in random matrices, Phys. Rev.
Lett. \textbf{75} , 69–72 (1995)




\bibitem{DKMVZ:99} Deift, P., Kriecherbauer, T., McLaughlin, K., Venakides,
S., Zhou, X.: Uniform asymptotics for polynomials orthogonal with
respect to varying exponential weights and applications to
universality questions in random matrix theory. Commun. Pure
Appl. Math. \textbf{52}, 1335-1425 (1999)


\bibitem{De-G:07} Deift, P., Gioev, D.: Universality in random matrix
theory for orthogonal and symplectic ensembles.
Int. Math. Res. Papers., 004-116 (2007)


\bibitem{Fo-Leb:13} Forrester, P.J. and  Lebowitz, J. L.
Local Central Limit Theorem for Determinantal Point Processes. Arxive 1311.7126v3

\bibitem{Jo:98} Johansson, K.: On fluctuations of eigenvalues of random
Hermitian matrices. Duke Math. J. \textbf{91}, 151-204 (1998)


\bibitem{Me:91} Mehta, M.L.: Random Matrices. New York: Academic
Press (1991)

\bibitem{K:13} Kargin,V.: On Pfaffian random point fields, J. Stat. Phys. DOI 10.1007/s10955-013-
 (2013)

\bibitem{KS:10} Kriecherbauer,T., Shcherbina, M.:
Fluctuations of eigenvalues of matrix models and their applications.
preprint arxive: math-ph/1003.6121


\bibitem{PS:97} Pastur, L., Shcherbina, M.: Universality of the local
eigenvalue statistics for a class of unitary invariant random
matrix ensembles. J. Stat. Phys. \textbf{86}, 109-147
(1997)

\bibitem{PS:07} Pastur, L., Shcherbina, M.:
Bulk universality and related properties of Hermitian matrix models.
J.Stat.Phys. \textbf{130},  205-250 (2007)

\bibitem{PS:11} Pastur, L., Shcherbina, M.:
Eigenvalue Distribution of Large Random Matrices. AMS,
 634 pp., SURV/171 (2011)


\bibitem{S:09} Shcherbina, M.: On  Universality  for Orthogonal Ensembles of Random Matrices
Commun.Math.Phys. \textbf{285}, 957-974 (2009)


\bibitem{S:11} Shcherbina, M.: Orthogonal and symplectic matrix models: universality and other properties
Commun.Math.Phys. \textbf{307},  761-790 (2011)

\bibitem{S:13} Shcherbina, M. Fluctuations of linear eigenvalue statistics of $\beta$ matrix models in the multi-cut regime
   J.Stat.Phys., \textbf{151}, N 6 ,  1004-1034 (2013)

\bibitem{S:14}  Shcherbina, M. Change of variables as a method to study general $\beta$-models: bulk
  universality  J. Math. Phys. \textbf{55}, N4, 043504 (2014)

\bibitem{So:00}  Soshnikov, A. Gaussian fluctuation for the number of particles in Airy, Bessel, Sine,
and other determinantal random point fields, J. Stat. Phys. \textbf{100} , 491–522 (2000).

\bibitem{Tr-Wi:98} Tracy,  C.A., Widom, H.: Correlation
functions, cluster functions, and spacing distributions for random
matrices. J.Stat.Phys. \textbf{92}, 809-835 (1998)

\bibitem{Wi:99} Widom, H.: On the relations between orthogonal, symplectic
and unitary matrix models. J.Stat.Phys. \textbf{94}, 347-363 (1999)
\end{small}
\end{thebibliography}
\end{document}